\documentclass[pdflatex,sn-basic]{sn-jnl}
\usepackage{graphicx}%
\usepackage{multirow}%
\usepackage{amsmath,amssymb,amsfonts}%
\usepackage{amsthm}%
\usepackage{mathrsfs}%
\usepackage[title]{appendix}%
\usepackage{xcolor}%
\usepackage{textcomp}%
\usepackage{manyfoot}%
\usepackage{booktabs}%
\usepackage{algorithm}%
\usepackage{algorithmicx}%
\usepackage{algpseudocode}%

\theoremstyle{thmstyleone}%
\theoremstyle{thmstyletwo}%
\newtheorem{remark}{Remark}%

\theoremstyle{thmstylethree}%

\begin{document}

\title[Copula-Based Clustering of Financial Time Series via Evidence Accumulation]{Copula-Based Clustering of Financial Time Series via Evidence Accumulation}

\author[1,2]{Andrea Mecchina}

\author[2]{\fnm{Roberta} \sur{Pappad\`{a}}}

\author[2]{\fnm{Nicola} \sur{Torelli}}

\affil[1]{\orgdiv{Department of Mathematics, Informatics and Geosciences}, \orgname{University of Trieste}, \orgaddress{\street{Via Alfonso Valerio 4/1}, \city{Trieste}, \country{Italy}}, \url{andrea.mecchina@phd.units.it}}

\affil[2]{\orgdiv{Department of Economics, Business, Mathematics and Statistics ``Bruno de Finetti''}, \orgname{University of Trieste}, \orgaddress{\street{Via Alfonso Valerio 4/1}, \city{Trieste}, \country{Italy}}, \url{rpappada@units.it}, \url{nicola.torelli@deams.units.it}}

\abstract{Understanding the dependence structure of asset returns is fundamental in risk assessment and is particularly relevant in a portfolio diversification strategy. We propose a clustering approach where evidence accumulated in a multiplicity of classifications is achieved using classical hierarchical procedures and multiple copula-based dissimilarity measures. Assets that are grouped in the same cluster are such that their stochastic behavior is similar during risky scenarios, and risk-averse investors could exploit this information to build a risk-diversified portfolio. An empirical demonstration of such a strategy is presented by using data from the EURO STOXX 50 index.}


\maketitle

\section{Introduction}\label{sec1}

Time series clustering includes a wide range of unsupervised learning techniques used to extract meaningful information in a variety of application fields. The clustering algorithm used to identify groups of time series may require the definition of a suitable distance or dissimilarity measure between any two time series. From a different perspective, a model-based setting assumes that the time series are generated by a stochastic process, and the clustering procedure aims to identify the time series that exhibit similar patterns. For some examples, see, for example, \cite{maharaj2019time, d2013autoregressive, liao2005clustering, piccolo1990distance}.

In some contexts, such as finance and environmental science, it is widely recognized that similarity measures that allow us to detect the comovements of time series can be more suitable to identify groups of time series with similar stochastic behavior, which may be relevant for risk assessment. In finance, the dependence among asset returns is crucial for portfolio selection and diversification. These methods originally made use of linear correlation; see, for instance, \cite{mantegna1999hierarchical} and \cite{bonanno2004networks}. In recent literature, alternative measures have been explored to quantify the pairwise association among time series based on other measures of association, concordance, or tail dependence, exploiting the potential of a copula-based approach; see, e.g., \cite{de2021hierarchical, fuchs2021dissimilarity, di2017copula}, and the references therein. 

In particular, when the interest is in modeling the dependence among a pair of time series by focusing on the left (or right) tail of their joint distribution, a natural approach is to define a dissimilarity measure as a function of tail dependence coefficients as done, for instance, in \cite{benevento2024tail, d2023tail, durante2014clustering, de2011tail}.
Tail dependence coefficients only depend on the copula function of two continuous random variables and have been extensively studied for many popular copula families; see, among others, \cite{schmid2010copula, charpentier2007lower, charpentier2009tails, gudendorf2010extreme, durante2015copulas}. 

However, these approaches are based on an unknown copula function $C$ that describes the dependence among a pair of time series. The selection of a specific copula model may be too restrictive and highly influences the resulting dissimilarity measure and, as a consequence, clusters identification. Moreover, while tail dependence coefficients describe extremal dependence at the asymptotic level, in some contexts, tail dependence in subasymptotic regimes may be more suitable to capture joint extreme values (see, e.g., \cite{durante2015copulas, sweeting2013calculating, patton2012review}). The choice of the quantile level used to measure tail association can lead to multiple risk scenarios. In addition, any set of identified clusters strongly depends on choices that are specific to the adopted clustering algorithm.

In order to overcome these limitations, we advocate clustering via evidence accumulation, as described in \cite{fred2002data}, i.e., we build an ensemble from multiple partitions, adopting alternative copula-based dissimilarities and clustering algorithms, to address tail dependencies among time series. The proposed approach allows us to achieve more robust results compared to those obtained by running a single clustering algorithm with a specified copula model; its practical relevance in the context of portfolio diversification is also discussed.

The paper is organized as follows. Section \ref{sec2} presents the framework and the chosen dissimilarity measure. Section \ref{sec3} illustrates the proposed clustering procedure. An application to the analysis of the components of the EURO STOXX 50 index is given in Sect. \ref{sec4}. Finally, Sect. \ref{sec5} concludes.

\section{Tail dependence measures for time series}\label{sec2}

As mentioned in Sect. \ref{sec1}, our aim is to focus on the dependence in the tail at the pairwise level, by means of a copula approach. We recall that copula functions offer a flexible way to specify a probability model for the dependent variables in the random vector $\mathbf{X}=(X_1, \dots, X_d)^T$, see \cite{durante2016principles} for an introduction to copulas. 
Copulas allow us to represent the joint distribution function $F$ of $\mathbf{X}$ in terms of its margins $F_i$, where $i = 1, \dots, d$ indexes the $d$ time series considered, and the associated copula $C$, which is unique for continuous margins \citep{sklar1959fonctions}:
\begin{equation}
F(x_1, \dots, x_d) = C(F_1(x_1), \dots, F_d(x_d)).\label{eq1}
\end{equation}
A $d$-dimensional copula $C(u_1,\dots, u_d)$ is simply a joint distribution over $[0,1]^d$ with uniform margins $u_i = F_i(x_i)$. Hence, the copula $C$ of the random vector $\mathbf{X}$ is a function mapping the univariate margins to the joint distribution $F$.

Our attention is restricted to the analysis of pairs of continuous variables, which in this case are time series. We will assume that our interest is in the lower tail of the join distribution, but the approach can be easily adapted to the case of upper tail dependence. The so-called tail dependence coefficients (TDCs) were originally suggested in \cite{sibuya1960bivariate}, and their copula-based representation was made explicit in \cite{joe1993parametric}. If limits exist, the lower TDC of the time series pair $(X_i, X_j)$, with also $j = 1, \dots, d$, is defined as
\begin{equation}
\lambda_{i | j}^L(C_{ij}) = \lim_{q \to 0^+} P(U_i < q | U_j < q) = \lim_{q \to 0^+}\frac{C_{ij}(q, q)}{q},\label{eq2}
\end{equation}
where $U_i$ and $U_j$ denote the probability integral transform of the random variables $X_{i}$ and $X_{j}$, respectively, i.e., $U_i = F_{i}(X_i)$ and $U_j =F_{j}(X_j)$. 
When the limit in Eq. \eqref{eq2} is greater than zero,  the components of $(X_i,X_j)$ are said to be asymptotically dependent in the lower tail, and when the limit is zero, they are said to be asymptotically independent.
Now, it is interesting to note that, for a level $q \in (0, 1/2]$, the quantity
\begin{equation}
\lambda_{i | j}^L(C_{ij}, q)=\frac{C_{ij}(q, q)}{q},\label{eq3}
\end{equation}
quantifies the extremal dependence for the pair $(i,j)$ at a finite level, and we will call it finite lower TDC. 
Thus, without relying on asymptotic dependence, Eq. \eqref{eq3} can be used to quantify the association between values in the lower left quadrant of the bivariate distribution of the pair indexed by $i$ and $j$, depending on the choice of a threshold level $q$. In finance, the interest is in assessing the risk of simultaneous losses on multiple assets. Then, assuming that a copula may describe the dependence between the components of a portfolio, different values of $q$ can be adopted to define risk scenarios based on the joint distribution, that is, the copula, of all pairs of logarithmic returns on the available assets.

\begin{remark}
When the interest is in the upper tail of the joint distribution of a pair of random variables $(X_i, X_j)$, the upper TDC is defined, if the limit exists, as
$\lambda_{i | j}^U(C_{ij}) = \lim_{q \to 1^-} \lambda_{i | j}^U(C_{ij}, q)$, where $\lambda_{i | j}^U(C_{ij}, q)= P(U_i > q | U_j > q) =(1 - 2q + C_{ij}(q, q))/(1 - q)$, for $q \in (1/2, 1)$.
\end{remark}

Let $\{X_{1t}\}, \dots, \{X_{dt}\}$ denote $d$ time series, with $t = 1, \dots, T$. Our aim is to model the dependence between each pair $(X_{it}, X_{jt})$, for $i, j = 1, \dots, d$, thereby taking into account the interdependence among the time series. As discussed in \cite{patton2012review}, copulas can be used in time series modeling to describe the conditional dependence of a random vector, given some information about its past behavior; see also \cite{patton2009copula} for an earlier review of copula modeling of financial time series. 

When considering copula-based models for multivariate time series, individual series are typically modeled first to tackle serial dependence or potential time-varying conditional mean and variance, and then dependence is modeled through the copula of the innovations. Following this idea, we assume that each time series $\{X_{it}\}$, for $i = 1, \dots, d$, can be modeled by a process of the form
\begin{equation}
X_{it} = \mu_i(\boldsymbol{Z}_{t - 1}) + \sigma_i(\boldsymbol{Z}_{t - 1}) \varepsilon_{it},\label{eq6}
\end{equation}
where the random variables $\boldsymbol{Z}_{t - 1}$ depend on $\mathcal{F}_{t - 1}$, the available information up to time $t - 1$ and the innovation series, $\varepsilon_{it}$, has a conditional distribution $F_{it}$ with mean $0$ and variance $1$, e.g. $\varepsilon_{it}$ given $\mathcal{F}_{t - 1}$ follows the distribution $F_{it}$.  The conditional mean and variance of each time series, namely $\mu_i$ and $\sigma_i$, are estimated by using a suitable parametric specification that allows for potentially time-varying conditional mean and variance, such as models belonging to the class of ARMA-GARCH models and their variants, as detailed in \cite{tsay2005analysis}. When the distribution of the innovation series, $F_{it}$, is modeled in a nonparametric way, it is assumed constant, that is $F_{it} = F_i$, for all $t$. Due to the results in \cite{chen2006estimation, remillard2017goodness}, the dependence among the original time series can be described in terms of the copula $C$ of the  standardized residuals from the marginal models; hence, under correct marginal specifications, rank-based copula inference procedures can be applied on the cross-sectionally dependent series $(\varepsilon_{1t},\dots, \varepsilon_{dt})$, indexed by $i = 1, \dots, d$ as the original series.

Based on previous considerations, our procedure to estimate the tail dependencies among the $d$ time series $\{X_{it}\}$ can be summarized as follows:
\begin{enumerate}
\item We fit an appropriate model of the form of Eq. \eqref{eq6} to each univariate time series to describe serial correlation, eventually allowing for models that can describe time-varying conditional mean and variance; the choice of these models is guided by classical model selection criteria, e.g., Akaike information criterion, and goodness-of-fit tests based on model residuals.
\item Using the parametric model estimated in the previous step, denoting by $\hat{\mu}_i$ and $\hat{\sigma}_i$ the fitted mean and volatility processes, we construct the time series of the estimated standardized residuals $\{\hat{\varepsilon}_{it}\}$, for $t = 1, \dots, T$ and each $i = 1, \dots, d$, as
\begin{equation}
\hat{\varepsilon}_{it} = \frac{x_{it} - \hat{\mu}_i(\boldsymbol{Z}_{t - 1})}{\hat{\sigma}_i(\boldsymbol{Z}_{t - 1})}\label{eq7}
\end{equation}
and obtain the so called \emph{pseudo-observations} time series $\{u_{it}\}$ via the estimated probability integral transform variables, $\hat{U}_{it}=\hat{F}_i(\hat{\varepsilon}_{it})$, where $\hat{F}_i$ denotes  the empirical distribution function of the $i$-th time series and 
$\{\hat{\varepsilon}_{it}\}$ the $i$-th time series of estimated standardized residuals.
\item For each pair of univariate time series we can estimate the underlying non-time-varying copula $C_{ij}$ of the series $\{\varepsilon_{it}\}$ and $\{\varepsilon_{jt}\}$ in the spirit of the inference functions for margins estimator (see \cite{joe1997multivariate, joe1996estimation}). The latter is a two-stage procedure that first involves estimating the marginal distributions $F_1, \dots, F_d$ in a nonparametric fashion. Then, the copula parameters are obtained by maximizing a log-likelihood-like function of a specified copula model, where the i.i.d. sample pair consisting of $U_{it} = F_i(\varepsilon_{it})$ and $U_{jt} = F_j(\varepsilon_{jt})$ is replaced by the sample pair of pseudo observations obtained as $\hat{U}_{it} = \hat{F}_i(\hat{\varepsilon}_{it})$ and $\hat{U}_{jt} = \hat{F}_j(\varepsilon_{jt})$ respectively, for $t\in \{1, \dots, T\}$. Finally, 
for any fixed level $q \in (0, 1/2]$, the coefficients in Eq. \eqref{eq3} can be obtained by replacing $C_{ij}$ with the corresponding bivariate estimated copula for the $i$-th and $j$-th time series.
\end{enumerate}

\subsection{Defining a dissimilarity measure}\label{subsec21}

A fundamental step in dissimilarity-based clustering algorithms is to obtain a suitable measure of similarity between each pair of objects to be clustered. In the context of copula-based measures, such a measure can be defined for each pair of continuous random variables, with copula $C$, as a mapping $\delta$ that is law-invariant, symmetric in its arguments, and assumes the value $0$ whenever a pair is coupled by the comonotonicity copula $M = \min(u, v)$, for $u$ and $v$ uniformly distributed over the unit interval. As discussed in \cite{fuchs2021dissimilarity, durante2024dissimilarity}, this mapping can be equivalently expressed as a functional from the space of bivariate copulas to $[0, +\infty)$.

Here, the tail association is of interest and we employ as a measure of pairwise dissimilarity between two time series $\{X_{it}\}$, $\{X_{jt}\}$ a suitable transformation of tail dependence coefficients, as done, for instance, in \cite{d2023tail, de2021hierarchical, de2011tail, durante2015clustering}. Thus, for each pair of time series $\{X_{it}\}$ and $\{X_{jt}\}$, we express the pairwise dissimilarity as
\begin{equation}
\delta_{ij}(C_{ij}, q) = \sqrt{2(1 - \lambda_{i | j}^L(C_{ij}, q))},\label{eq8}
\end{equation}
where $\lambda_{i | j}^L(C_{ij}, q)$ is defined as in Eq. \eqref{eq3} and only depends on the bivariate copula $C_{ij}$ and a fixed quantile $q \in (0,1/2]$. Notice that the above measure gives values of dissimilarities ranging from $0$, when the tail dependence is maximum, to $\sqrt{2}$, when the tail dependence is null. Thus, the dissimilarity measure in Eq. \eqref{eq8} captures the strength of the dependence in the lower tail between the two variables compared. As a result, the final clusters will group time series with similar behavior under risky scenarios, i.e., when simultaneous losses occur.

\begin{remark}
In the literature, both parametric estimators, e.g., see \cite{de2011tail}) and nonparametric ones, e.g., see \cite{durante2015clustering}), have been considered for the pairwise tail dependence coefficients in the clustering framework. It is worth noting that the tail behavior of a multivariate model depends solely on the specific copula model and not on the marginal distributions. Hence, in modeling the dependencies of extremes, the choice of a copula family plays a crucial role.
\end{remark}

The pairwise dissimilarities in Eq. \eqref{eq8} are used to build a $d\times d$ dissimilarity matrix, denoted as $\Delta$, which requires the estimation of $d(d-1)/2$ bivariate copulas. 
Such matrix could be used as input for dissimilarity-based clustering algorithms, such as agglomerative hierarchical clustering techniques, see \cite{everitt2011hierarchical}. However, we propose an alternative approach to cluster the $d$ time series that takes advantage of multiple partitions arising from several copula models and quantile levels to enhance the clustering results. This procedure can be referred to as an ensemble-based clustering algorithm and is described in detail in the next section.

\section{Ensemble-based clustering algorithm}\label{sec3}

Let $\mathcal{C}$ be a set of $n$ bivariate parametric copula families, e.g., elliptical and Archimedean copulas, which describe different behaviors in the upper and lower tail of a bivariate distribution. Let $q \in \mathcal{Q}$, where $\mathcal{Q}$ is a discrete set of $m$ values in the interval $(0, 1/2]$. Starting from the $d$ time series $\{X_{it}\}$, for $i\in\{1, \dots, d\}$, one can obtain multiple dissimilarity matrices from Eq. \eqref{eq8}, by considering  combinations of elements in $\mathcal{C}$ and $\mathcal{Q}$.  Specifically, the proposed procedure is as follows.
\begin{enumerate}
\item For each pair $(i,j)$ of time series, each copula family $C \in \mathcal{C}$ and each $q \in \mathcal{Q}$, estimate the extremal dependence coefficient $\hat{\lambda}_{i|j}^L$ according to Eq. \eqref{eq3}, as illustrated in Sect. \ref{sec2}. 
\item Build $nm$ dissimilarity matrices $\Delta(C, q)$, $\forall C \in \mathcal{C}$ and $\forall q \in \mathcal{Q}$, such that the entries are  
defined as in Eq. \eqref{eq8} via a function of the estimated coefficients obtained in the previous step.
\item An ensemble of partitions is then obtained by applying any dissimilarity-based clustering method to each dissimilarity matrix $\Delta$ obtained in Step 2. Here, hierarchical agglomerative clustering is adopted, which produces a \emph{dendrogram} from which a final partition can be derived by cutting this tree at a suitable height.
\end{enumerate}

Step 3 of the above procedure requires some crucial user's choices. It is worth to recall that  hierarchical clustering methods require a specific rule for merging two clusters called \emph{linkage}, specifying how the dissimilarity between two clusters is computed as a function of the pairwise dissimilarities between the clusters. The most used linkages are the single, the average, and the complete linkage; see \cite{everitt2011hierarchical}. 

The final result of the clustering procedure could be influenced by the choice of the linkage, therefore we decide to include in the ensemble the partitions obtained by applying the average and complete linkage. We do not consider the single linkage because of its well known tendency to produce singletons. Finally, to obtain the  partition, an appropriate number of clusters must be selected. Visual examination of the dendrogram and classical validation indices are often combined to identify an appropriate number of clusters $k^{\star}$ within an appropriate range $\{k_{\min}, \dots, k_{\max}\}$.

We note that a dissimilarity matrix could be approximated by a Euclidean distance matrix, e.g., by using multidimensional scaling so that the interpoint distances closely match the input dissimilarities. Then, the obtained point configurations can be used as input for the classical $k$-means algorithm, as discussed in \cite{durante2014clustering}. This method is faster than hierarchical clustering, but the number of clusters must be fixed in advance.

\subsection{Evidence accumulation clustering}\label{subsec22}

With the aim to exploit the information contained in multiples partitions, we adopt the idea of evidence accumulation for combining the results of various  clusterings into a single data partition, by viewing each distinct clustering result as an independent evidence of data organization \citep{fred2002data}. Specifically, a cluster ensemble is used to obtain the occurrences of time series pairs in the same cluster across the different partitions. The underlying assumption is that patterns belonging to a genuine cluster are more likely to be located in the same cluster across different partitions.

Each partition of the $d$ time series into $k$ groups is generated by a clustering algorithm that requires as input a copula model $C \in \mathcal{C}$, the quantile level $q \in \mathcal{Q}$, and the linkage $l\in \mathcal{L}$, with $|\mathcal{L}|=r$. Hence, the ensemble of partitions has dimension $nmr$.

Let $h_i(C, q, l)$ denote the cluster label of the $i$-th time series in the partition of the ensemble associated with the copula model $C \in \mathcal{C}$, the quantile level $q \in \mathcal{Q}$, and the linkage $l$. For each pair of time series $(i,j)$, we define a partition vote as
\begin{equation}
v_{ij}(C, q, l) = 
\begin{cases}
1 & \text{if } h_i(C, q, l) = h_j(C, q, l) ,\\
0 & \text{otherwise}.
\end{cases}\label{eq9}
\end{equation}
Thus, $v_{ij}(C, q, l)$ simply expresses whether the time series $i$ and $j$ are assigned to the same partition, for a given combination of $C, q, $ and $l$. Taking the co-occurrences of pairs of time series in the same cluster as votes for their association, the data partitions in the cluster ensemble are mapped into a symmetric co-association matrix $V$ of size $d\times d$, with elements
\begin{equation}
V_{ij}= \sum_{C \in \mathcal{C}}\sum_{q \in \mathcal{Q}}\sum_{l \in \mathcal{L}}v_{ij}(C, q, l).
\label{eq10}
\end{equation}
Recalling that $nmr$ clusterings are available, such co-association matrix is converted to a consensus matrix $M$, whose entries are the proportions of partitions in the ensemble in which the time series indexed by $i$ and $j$ are located together, namely:
\begin{equation}
M_{ij}= \frac{V_{ij}}{nmr}.\label{eq11}
\end{equation}
Clearly, $M_{ij} \in [0, 1]$, and $M_{ii} = 1$ for all $i, j \in {1, \dots, d}$. If the items in the matrix were arranged so that items belonging to the same cluster are adjacent to each other, perfect consensus would translate into a block-diagonal matrix with non-overlapping blocks of ones along the diagonal, with each block corresponding to a different cluster, surrounded by zeros \citep{monti2003consensus}.

In order to derive a final clustering into $k$ clusters based on the proposed consensus approach, the consensus matrix $M$ is converted into a dissimilarity matrix $D$, whose elements are obtained as
\begin{equation}
D_{ij}= 1 - M_{ij}.\label{eq12}
\end{equation}
Finally, we apply agglomerative hierarchical clustering to the symmetric $d \times d$ matrix $D$ of elements $D_{ij}$ to obtain a final partition, with a specified linkage method. We note that, at this final stage, the choice of the linkage method is expected to be negligible, as it has already been included in the process for building the ensemble.

\section{Application to financial time series}\label{sec4}

In order to illustrate our approach, we analyze daily log-returns of the components of the EURO STOXX 50 index in the period from January 1, 2018 to December 31, 2022. As out-of-sample period, we consider the period from July 1, 2023 to September 30, 2023. Such a period has been selected due to the fact that EURO STOXX 50 was experiencing severe losses, as further discussed in Sect. \ref{sec4}, which are of utmost importance for the considered application. The analyses are carried out by restricting our attention to the assets included in the index for the whole training and testing periods, relying on information sourced from the index provider STOXX Ltd.\footnotemark[1], leading to a dataset composed of $d = 38$ asset time series. By considering only the days when all assets were operating, i.e., by excluding weekend days and national holidays, we collect $d=38$ time series of length $T = 1195$  for the training data set; Table \ref{tab1} lists the selected assets, as well as their tickers, supersectors and countries.
\footnotetext[1]{Source: \href{https://stoxx.com/index/sx5e/}{https://stoxx.com/index/sx5e/}, with a particular attention the ``DATA" and ``ANNOUNCEMENTS" sections for the components and their updates, accessed on 8 August, 2025.}

\begin{table}[h]
\caption{Selected assets from the EURO STOXX 50 index}\label{tab1}%
\begin{tabular}{@{}p{0.25\linewidth}lll@{}}
\toprule
Asset name & Ticker symbol & Supersector & Country \\
\midrule
Adidas & ADS.DE & Sportwear & Germany \\
Ahold Delhaize & AD.AS & Retail & Netherlands \\
Air Liquide & AI.PA & Chemistry & France \\
Airbus & AIR.PA & Aerospace & France \\
Allianz & ALV.DE & Insurance & Germany \\
Anheuser-Busch InBev & ABI.BR & Food and beverage & Belgium \\
ASML Holding & ASML.AS & Technology & Netherlands \\
AXA & CS.PA & Insurance & France \\
BASF & BAS.DE & Chemistry & Germany \\
Bayer & BAYN.DE & Chemistry & Germany \\
Banco Santander & SAN.MC & Banking & Spain \\
BMW & BMW.DE & Automotive industry & Germany \\
BNP Paribas & BNP.PA & Banking & France \\
Danone & BN.PA & Food and beverage & France \\
Deutsche Post & DHL.DE & Logistics & Germany \\
Deutsche Telekom & DTE.DE & Telecommunication & Germany \\
Enel & ENEL.MI & Electric utility & Italy \\
Eni & ENI.MI & Petroleum & Italy \\
EssilorLuxottica & EL.PA & Optical industry & France \\
Flutter Entertainment & FLTR.L & Bookmaking & Ireland \\
Herm\`{e}s & RMS.PA & Luxury & France \\
Iberdrola & IBE.MC & Electric utility & Spain \\
Inditex & ITX.MC & Clothing & Spain \\
Infineon Technologies & IFX.DE & Semiconductors & Germany \\
ING Group & INGA.AS & Banking & Netherlands \\
Intesa Sanpaolo & ISP.MI & Banking & Italy \\
L'Or\'{e}al & OR.PA & Personal and household goods & France \\
LVMH Mo\"{e}t Hennessy Louis Vuitton & MC.PA & Personal and household goods & France \\
Mercedes-Benz Group & MBG.DE & Automotive industry & Germany \\
Munich Re & MUV2.DE & Insurance & Germany \\
Safran & SAF.PA & Aerospace & France \\
Sanofi & SAN.PA & Pharmaceutical industry & France \\
SAP & SAP.DE & Technology & Germany \\
Schneider Electric & SU.PA & Goods and Services & France \\
Siemens & SIE.DE & Goods and Services & Germany \\
TotalEnergies & TTE.PA & Petroleum & France \\
Vinci SA & DG.PA & Construction and materials & France \\
Volkswagen Group & VOW.DE & Automotive industry & Germany \\
\botrule
\end{tabular}
\footnotetext{Source: data were collected via the Yahoo Finance API using the \texttt{yfinance} Python package, see \cite{aroussi2023reliably}, on 8 August, 2025.}
\end{table}

In order to remove autocorrelation and heteroscedasticity from the univariate time series, we preliminary fit ARMA-GJR-GARCH$(1, 1)$ models with Student's $t$ distributed errors to tackle time-varying volatility and heavy tails. The orders of the ARMA models are selected according to the Akaike information criterion. All the univariate marginal models provide an adequate fit based on the classical tests and diagnostic tools applied to the standardized residuals and squared standardized residuals. Appendix \ref{secA1} provides additional details on these models.

Following the semiparametric procedure outlined in Sect. \ref{sec2}, from the standardized residuals time series $\{\hat{\varepsilon}_{it}\}$, for $i = 1, \dots, d$ and $t = 1, \dots, T$, the uniform variables $\hat{U}_{it}=\hat{F}_i(\hat{\varepsilon}_{it})$ are obtained by applying the empirical distribution function $\hat{F}_i$ proper to each time series. Then, for all $d(d - 1) / 2 = 703$ pairs of such variables $(\hat{U}_{it}, \hat{U}_{jt})$, with $i, j = 1, \dots, d$ and $i\neq j$, the Python package \texttt{pycop}, introduced in \cite{nicolas2022pycop}, is used to fit by maximum likelihood estimation $n = 8$ bivariate copula models. Two elliptical copulas are considered, the Gaussian copula ($C_1$) and the Student's $t$ copula $(C_2)$, as well as several Archimedean copulas, namely: the Clayton copula ($C_3$), the survival Gumbel copula ($C_4$), the Frank copula ($C_5$), the survival Joe copula ($C_6$), the survival Galambos copula ($C_7$) and the BB1 copula ($C_8$). 

\begin{figure}[h]
\centering
\includegraphics[width=.7\linewidth]{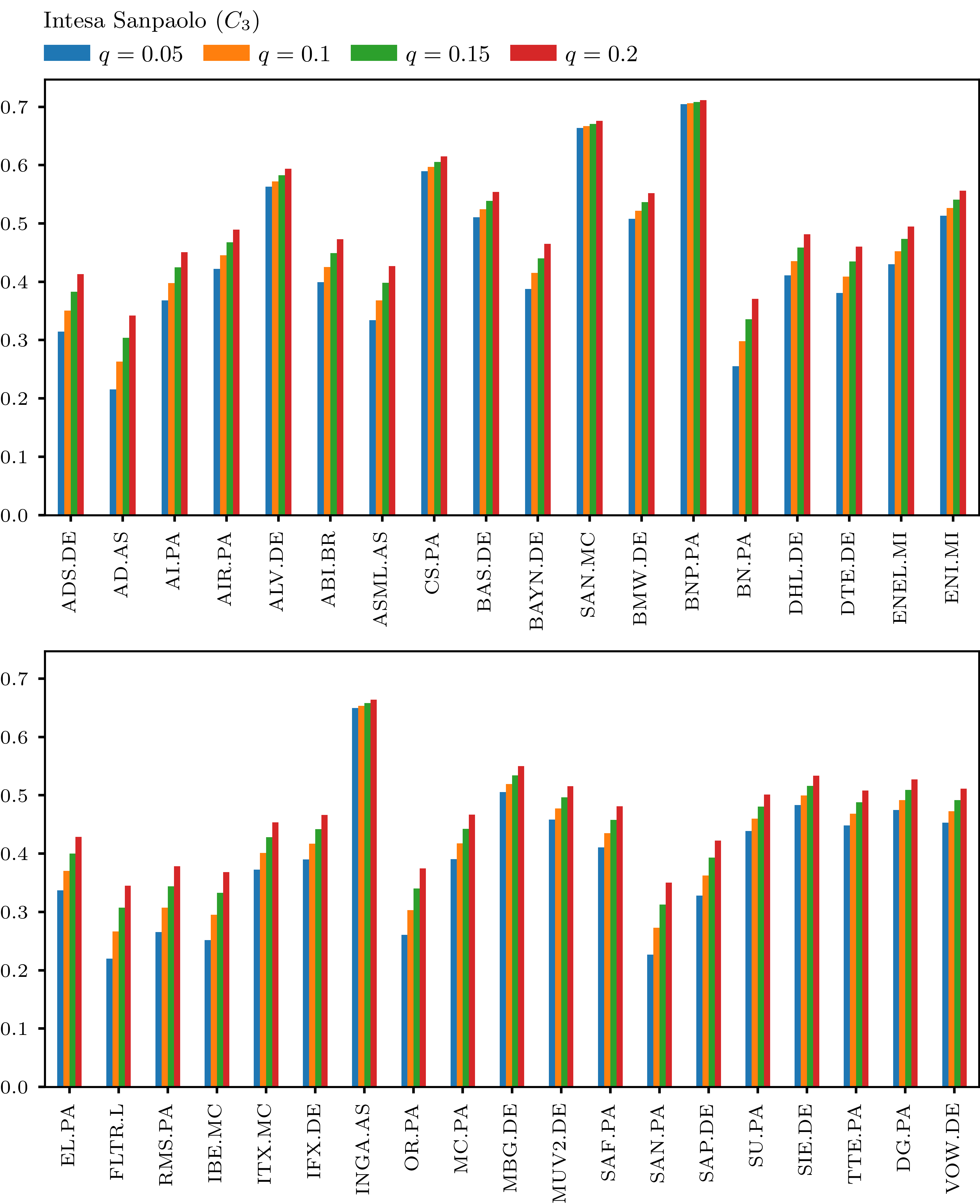}
\caption{Finite lower TDCs relative to the Intesa Sanpaolo asset under the Clayton copula specification, for the $m = 4$ different threshold values in $\mathcal{Q}$. Ticker symbols are used as labels for brevity; we refer to Table \ref{tab1} for the full correspondences.}\label{fig1}
\end{figure}

\begin{figure}[h]
\centering
\includegraphics[width=.7\linewidth]{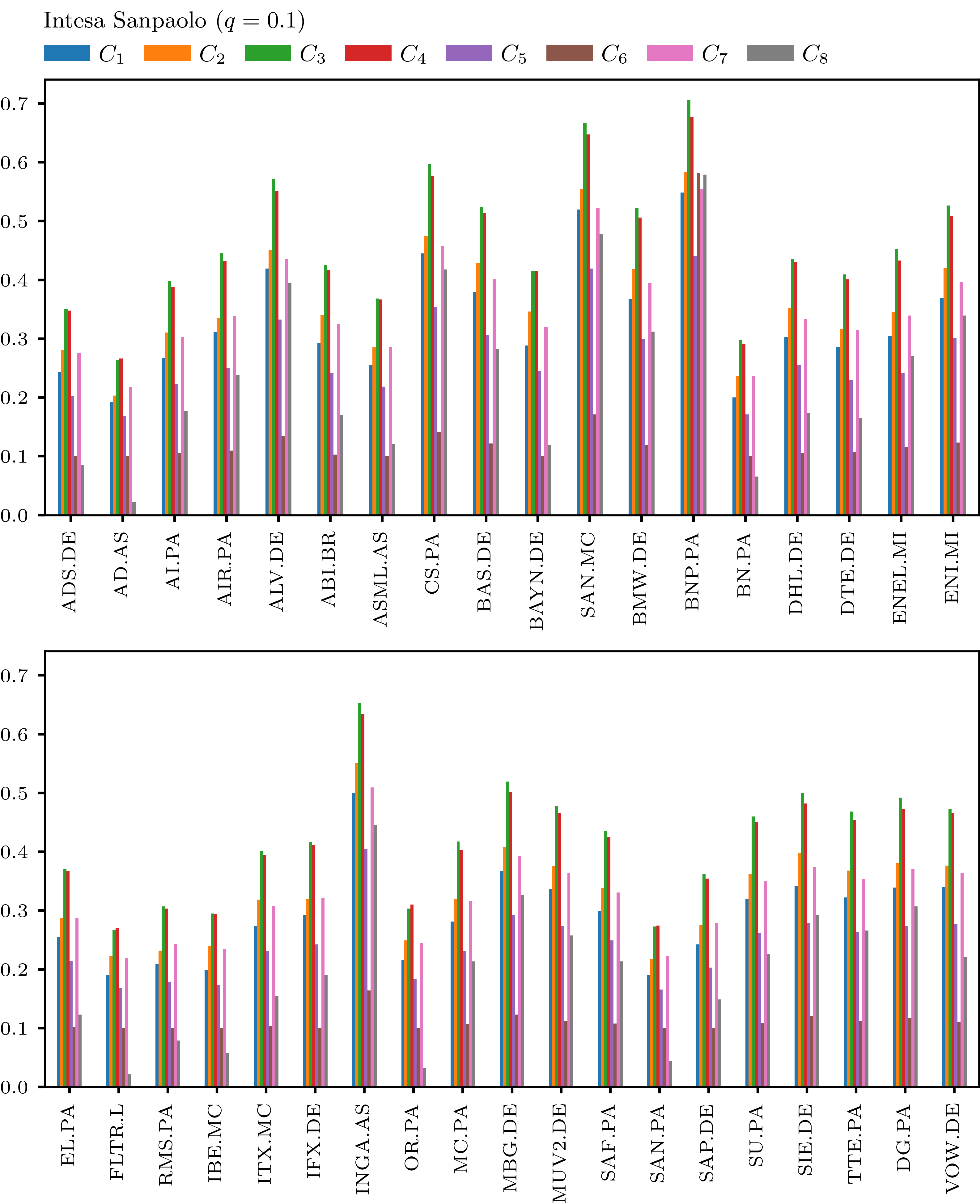}
\caption{Finite lower TDCs relative to the Intesa Sanpaolo asset for the threshold $q = 0.1$, under the $n = 8$ different of copula model specifications $C_1, \dots, C_8$. Ticker symbols are used as labels for brevity; we refer to Table \ref{tab1} for the full correspondences.}\label{fig2}
\end{figure}

We choose $m = 4$ low threshold values by specifying $\mathcal{Q} = \{0.05, 0.1, 0.15, 0.2\}$ and we evaluate the finite lower TDCs defined in Eq. \eqref{eq3}. An representative example of the results obtained is reported in Figure \ref{fig1}, which compares the finite lower TDCs relative to the Intesa Sanpaolo asset for the different values in $\mathcal{Q}$ under the Clayton copula. Conversely, Figure \ref{fig2} compares the finite lower TDCs obtained under the different copula specifications when $q = 0.1$ is kept fixed, for the same asset. Similar results are observed across different assets, copula models, and threshold values.

\begin{figure}[h]
\centering
\includegraphics[width=.7\linewidth]{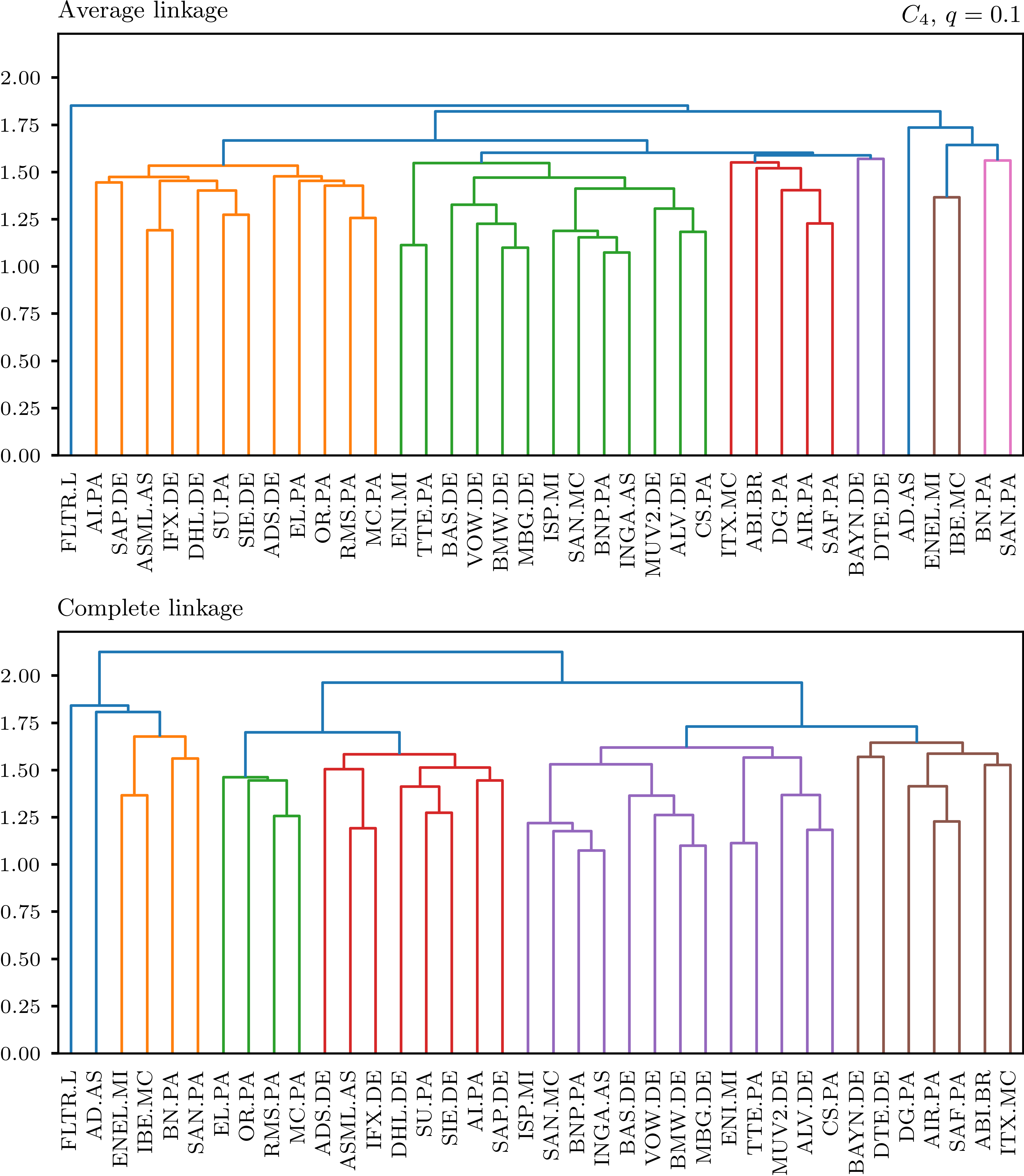}
\caption{Dendrograms obtained from the dissimilarity matrix relative to the survival Gumbel copula ($C_4$) with $q = 0.1$, respectively adopting the average (top) and the complete (bottom) linkage methods. The optimal number of clusters, respectively $8$ and $7$, maximizes the average silhouette score. The resulting partitions are highlighted with different colors; however, coloring is independent across the two dendrograms and does not imply any correspondence. The assets Flutter Entertainment and Ahold Delhaize each form a singleton cluster in both dendrograms. Ticker symbols are used as labels for brevity, see Table \ref{tab1}.}\label{fig3}
\end{figure}

The finite lower TDCs just found are stored in $nm = 32$ matrices of shape $d \times d$, which are in turn converted into dissimilarity matrices by Eq. \eqref{eq8}. Such matrices are used as input for agglomerative hierarchical clustering where two linkage methods are used, namely the average and complete linkages ($r=2$). The final partition is then obtained by selecting the number of clusters as the value maximizing the average silhouette score, see \cite{rousseeuw1987silhouettes}. Here, we set $k_{\min} = 5$ and $k_{\max} = 10$, to avoid solutions with too few clusters or clusters with very low cardinality.
Finally, we obtain an ensemble of $nmr = 64$ partitions. For illustration purposes, the dendrograms produced by the survival Gumbel copula ($C_4$), $q = 0.1$ and two different linkage methods are reported in Figure \ref{fig3}. The choice of the linkage method significantly affects the resulting partitions. Similar results are obtained across the different configurations in the ensemble.

\begin{figure}[h]
\centering
\includegraphics[width=.7\linewidth]{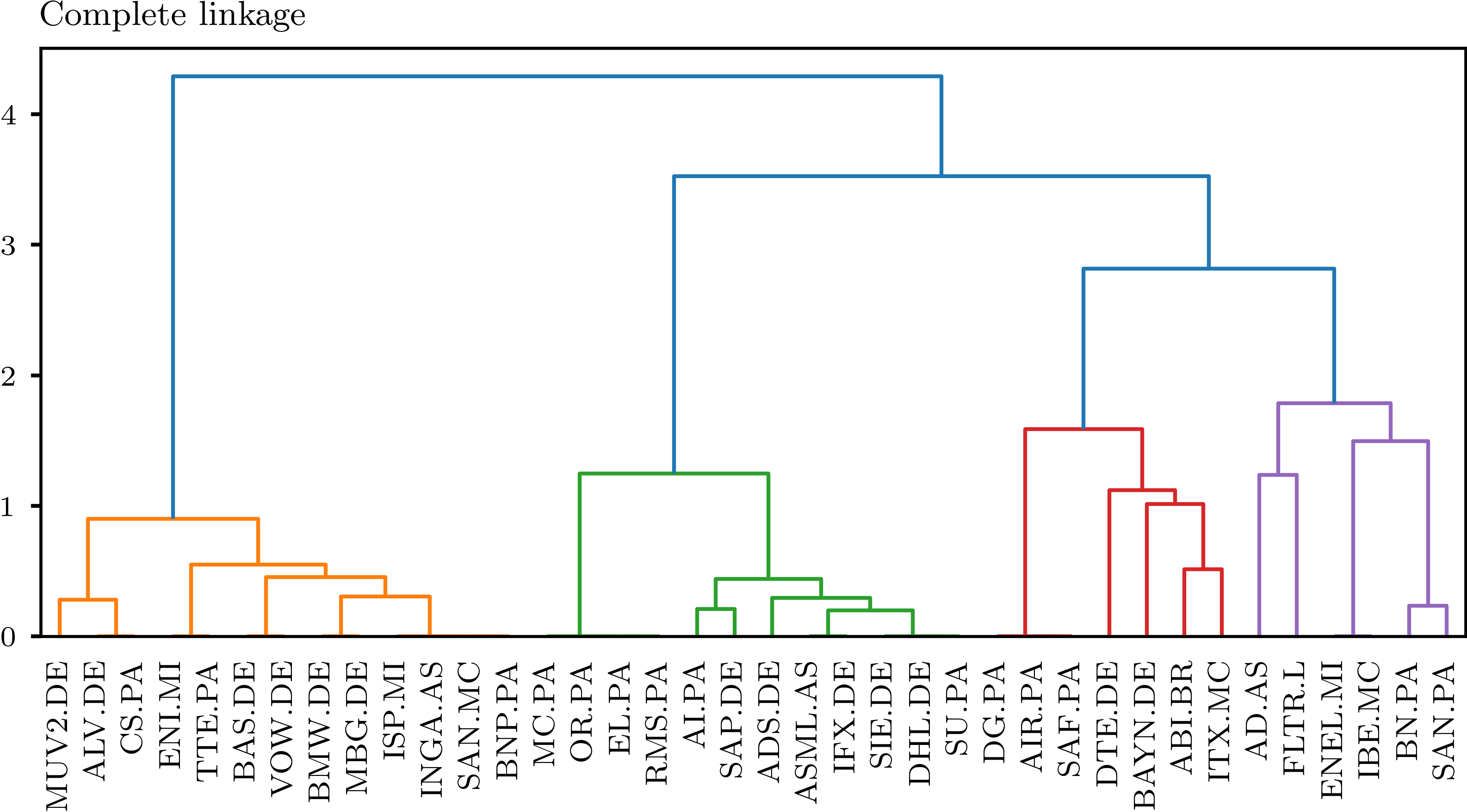}
\caption{Final dendrogram obtained from the dissimilarity matrix based on the co-occurrences of assets in the same partition. Distinct colors highlight the $4$ clusters obtained by maximizing the difference between subsequent merge distances. To obtain such figure, agglomerative hierarchical clustering is performed using the complete linkage method; however, the same clusters are obtained when using the average linkage method. Ticker symbols are used as labels for brevity, see Table \ref{tab1}.}
\label{fig4}
\end{figure}

As detailed in Sect. \ref{sec3}, after computing the co-occurrences of assets in the same cluster, a dissimilarity matrix is obtained as defined by Eq. \eqref{eq12}. The final partition is obtained by applying to such dissimilarity matrix the agglomerative hierarchical clustering algorithm once more. Several techniques can be used to cut the final dendrogram and obtain the resulting clustering, including visual inspection of the dendrogram itself, depending on the needs of the practitioner. The dendrogram we obtained is reported in Figure \ref{fig4}: the cut, leading to a final partition with $k^{*}=4$ clusters, maximizes the difference between subsequent merge distances, thus capturing a prominent natural division emerging from the data themselves. The flexibility and interpretability of such simple cut criterion is preferred, in this final step, to more complex and automated criteria like the maximization of the average silhouette score, as opposed to what is done for the single partitions in the ensemble. The reported dendrogram is obtained by applying the complete linkage method, but the same number of clusters is found with the average linkage method. The two linkage methods lead to two perfectly matching partitions, with an Adjusted Rand Index (ARI), defined in \cite{hubert1985comparing}, exactly equal to $1$. This effect may be attributed to the inclusion of the linkage method as an additional degree of freedom in the ensemble: as a result, the final partition exhibits robustness with respect to the choice of the linkage method. This indeed aligns with the fundamental objective of the evidence accumulation approach, namely ensuring robustness against potential misspecifications of the threshold parameter $q$ and, most critically, of the underlying copula model.

We report in Table \ref{tab2} the assets grouped by the different clusters, whose sizes are respectively $13$, $12$, $7$ and $6$; the assets are listed alphabetically within each cluster. We propose a characterization of the $4$ clusters in terms of the constituent assets:
\begin{itemize}
\item Cluster $1$: financials, industry, and energy. This cluster consists of financial institutions (Allianz, AXA, Banco Santandander, BNP Paribas, ING Group, Munich Re), automotive and industrial companies (BMW, Mercedes-Benz Group, Volkswagen Group) and energy/chemical firms (Eni, TotalEnergies, BASF);
\item Cluster $2$: luxury, technology, and logistics. This cluster comprises companies producing luxury and retail goods (Herm\`{e}s, LVMH Mo\"{e}t Hennessy Louis Vuitton, L'Or\'{e}al, EssilorLuxottica, as well as Adidas), technology companies (ASML Holding, Infineon Technologies, SAP) and smaller industries and logistic firms (Deutsche Post, Schneider Electric, Siemens, as well as Air Liquide);
\item Cluster $3$: aerospace, infrastructure, and consumer staples. This cluster mainly includes aerospace industries (Airbus, Safran), telecommunications and infrastructure companies (Deutsche Telekom, Vinci SA), as well as firms in the consumer staples (Anheuser-Busch InBev, Inditex) and pharmaceutical (Bayer) fields;
\item Cluster $4$: utilities, retail and healthcare. This cluster consists of utilities companies (Enel, Iberdrola), and firms operating in the retail and entertainment (Ahold Delhaize, Danone, Flutter Entertainment) and healthcare (Sanofi) fields.
\end{itemize}
We observe that the clusters found by the evidence accumulation strategy allow for a meaningful interpretation, offering deeper insights than a simple characterization by supersectors and countries, as might be inferred from Table \ref{tab1}. Further interpretations could be developed by inspecting the exposures to the different risks underlying each cluster of assets.

\begin{table}[ht!]
\caption{The $d = 38$ analyzed assets, grouped into the $4$ clusters identified by the evidence accumulation clustering strategy. For the sake of clarity, assets are reported by their full names}\label{tab2}%
{\footnotesize{
\begin{tabular}{llll}
\toprule
Cluster $1$ & Cluster $2$ & Cluster $3$ & Cluster $4$ \\
\midrule
Allianz & Adidas & Airbus & Ahold Delhaize \\
AXA & Air Liquide & Anheuser-Busch InBev & Danone \\
BASF & ASML Holding & Bayer & Enel \\
Banco Santander & Deutsche Post & Deutsche Telekom & Flutter Entert. \\
BMW & EssilorLuxottica & Inditex & Iberdrola \\
BNP Paribas & Herm\`{e}s & Safran & Sanofi \\
Eni & Infineon Technologies & Vinci SA & \\
ING Group & L'Or\'{e}al &  & \\
Intesa Sanpaolo & Louis Vuitton &  & \\
Mercedes-Benz Group & SAP &  & \\
Munich Re & Schneider Electric &  & \\
TotalEnergies & Siemens &  & \\
Volkswagen Group &  &  & \\
\botrule
\end{tabular}}}
\end{table}

\subsection{A portfolio diversification strategy}\label{subsec41}

As a way to further validate the asset partition we obtained, which we recall being reported in Table \ref{tab2}, in this Subsect. we explore the potential benefits our clustering offers with respect to several competitors in a portfolio diversification backtesting. The setting of portfolio diversification emerges as natural, considering that we are considering finite lower TDCs to model dependencies among the assets in a portfolio. If we focused on finite upper TDCs, discussed in the first Remark from Sect. \ref{sec2}, we should be testing an offensive portfolio strategy instead.

As anticipated at the beginning of Sect. \ref{sec4}, we consider as backtesting period the third quarter of 2023, from July 1, 2023 to September 30, 2023, which we identified as a turmoil period of $55$ trading days; indeed, the closing price of EURO STOXX 50 dropped from 4398.15 to 4174.66, realizing a log-return of $-5.22\%$. We consider a diversification strategy based on our clustering results analogous to the one proposed in \cite{wang2016portfolio}: among all possible portfolios composed of at most one asset per cluster, we pick the one which minimizes the conditional value-at-risk. Such measure, introduced in \cite{rockafellar2000optimization}, quantifies the expected loss realized by the portfolio in the worst $\alpha\%$ of cases; we set such parameter to $\alpha = 0.2$, in order to include a wider range of adverse market scenarios, from moderate stress to deep crisis. We point out that, despite $\alpha$ being conceptually similar to the quantile threshold $q$ of the proposed methodology, there is no constraint forcing us to set them to the same value. It is important to note that the primary objective of this portfolio strategy is not to present a computationally efficient and scalable portfolio diversification strategy, since all possible portfolio configurations are explored by brute force, but rather to highlight the economic value and the discriminatory power of our clustering methodology.

However, a direct comparison of the proposed method with the plain EURO STOXX 50 would not be methodologically fair, as our investible universe is restricted to the only $d = 38$ assets that were present in the index for the entire training and testing period considered.  For this reason, we consider several competitor portfolio strategies, all constructed restricting to the available $d = 38$ assets, that are: 
\begin{itemize}
\item the equally weighted (EW) portfolio, discussed in \cite{demiguel2009optimal};
\item the global minimum variance (GMV) portfolio, assuming a null risk-free rate, for which we refer to \cite{haugen1991efficient};
\item the portfolio which minimizes the global conditional value-at-risk at the $\alpha = 0.2$ risk level (min-$\text{CVaR}_{20\%}$);
\item two more clustering-based strategies, relying on Student's $t$ and BB1 copulas respectively, similar to what is done in \cite{d2023tail}.
\end{itemize}
The aforementioned clustering-based strategies model dependencies with two distinct copulas, which are used to obtain lower TDCs, in the limit of $q \to 0^+$ from Eq. \eqref{eq2}. Such coefficients are then plugged into our pipeline, leading to partitions for which only the linkage method (average or complete) is left as a degree of freedom. We find that simply cutting the dendrogram while maximizing the difference between subsequent merge distances tends to underestimate the number of cluster, which in this case would result $2$ or $3$, separating respectively $1$ or $2$ singletons. For this reason,  the optimal number of clusters is chosen by maximizing the average silhouette score within $k_{\min} = 5$ and $k_{\max} = 10$, analogously to what we did for the proposed clustering algorithm when dealing with dendrograms from single copulas and quantile thresholds. The selected number of clusters is $k^{*}=5$ for the Student's $t$ copula and  the BB1 copula  with average and complete linkage, respectively,  $k^{*}=10$ for the Student's $t$ copula and complete linkage, $k^{*}=8$ for the BB1 copula and average linkage. Portfolio weights are kept fixed, for the all the considered strategies, throughout the whole testing period.

We consider the following portfolio metrics to compare the performance of the considered diversification strategies on the considered backtest, namely: the expected return $\mu$, the volatility $\sigma$, the conditional value-at-risk at the $\alpha = 0.2$ risk level $\text{CVaR}_{20\%}$, the maximum drawdown (MDD), and the certainty equivalent (CE). Again, a null risk-free rate is assumed, as well as a unit risk-aversion coefficient. All the considered metrics are standard portfolio measures of performance, for which we refer to \cite{alexander2009market} or \cite{elton2009modern}. All measures refer to the whole $55$ trading days backtesting. The values we obtain are reported in Table \ref{tab3}; all the metrics, except for MDD, are annualized to ease their interpretability. The strategy based on the proposed evidence accumulation clustering method is indicated as `Ensemble''. For all metrics except expected return, lower values indicate better performance. The results for the clustering-based strategy relying on the Student's $t$ copula ($t$-copula) is reported only for the complete linkage method, as it uniformly outperforms the average linkage method; the converse holds for the BB1 copula (BB1-copula).

\begin{figure}[ht]
\centering
\includegraphics[width=.7\linewidth]{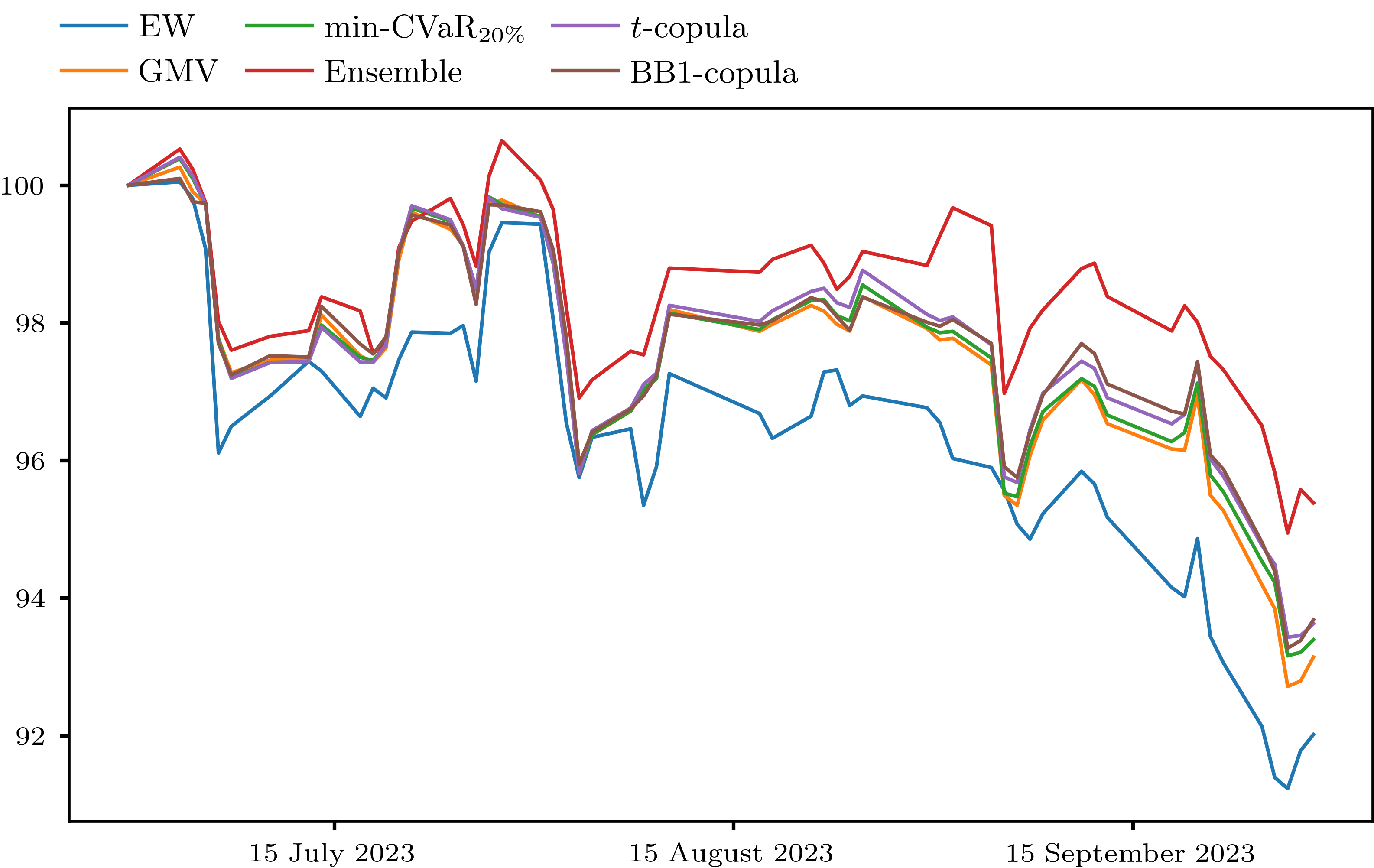}
\caption{Cumulative returns, re-based at $100$, for the several considered portfolio diversification strategies over the backtesting period covering the third quarter of 2023.}\label{fig5}
\end{figure}

Remarkably, the strategy based on the proposed clustering outperforms the competitors with respect to the all the considered metrics. We point out a lower value of $\text{CVaR}_{20\%}$ is obtained with respect to the strategy explicitly minimizing such quantity, because we benefit from the diversification effect of picking at most one asset per cluster, thus at most $4$ assets. Indeed, the min-$\text{CVaR}_{20\%}$ allocates sensible weights (larger than $1\%$) to $10$ assets. Finally, as already commented about the clustering-based approach relying on the Student's $t$ and BB1 copulas, we observe that the linkage method providing better performances is not the same for both copula models. Our proposed copula based strategy gets rid of this ambiguity by including the linkage method as a degree of freedom in the ensemble. The cumulative returns for the different strategies under consideration are reported in Figure \ref{fig5}. The curves, rebased at $100$, track the performance of a capital of $100$, invested on the first day and for the whole backtest period, according to each strategy. We observe that the cumulative return curve relative to our ensemble-based clustering strategy almost consistently dominates the others. It results in the smallest capital loss following the turbulent quarter backtested, demonstrating more successful risk diversification compared to all considered competitors.

\begin{table}[ht]
	\caption{Performance metrics (annualized except for MDD) for several diversification strategies, compared over the backtesting period covering the third quarter of 2023}\label{tab3}%
	\begin{tabular}{@{}llllll@{}}
		\toprule
		Diversification strategy & $\mu$ & $\sigma$ & $\text{CVaR}_{20\%}$ & MDD & CE \\
		\midrule
		EW & -32.52\% & 12.68\% & 20.28\% & 8.81\% & -33.32\% \\
		GMV & -28.53\% & 11.67\% & 19.49\% & 7.53\% & -29.21\% \\
		min-$\text{CVaR}_{20\%}$ & -27.61\% & 11.57\% & 19.31\% & 7.20\% & -28.28\% \\
		Ensemble & -20.02\% & 11.14\% & 17.18\% & 5.67\% & -20.64\% \\
		$t$-copula & -26.75\% & 11.65\% & 19.31\% & 6.95\% & -27.43\% \\
		BB1-copula & -26.54\% & 11.63\% & 19.29\% & 6.82\% & -27.21\% \\
		\botrule
	\end{tabular}
\end{table}

\section{Conclusions}\label{sec5}

In this contribution, we presented a clustering method that enables us to group time series according to pairwise comovements in the lower-left tail of their joint distribution. This method exploits evidence arising from (i) several copula models, (ii) different quantiles for the extremal lower tail dependence, and (iii) linkage methods for the agglomerative hierarchical clustering procedure. The final partition we obtain turns out to be robust to misspecifications of the copula model, of the quantile threshold, and of the linkage method as well. We note that the proposed strategy is, in general, viable in pair with any clustering algorithm that takes a dissimilarity matrix as input, simply by replacing the linkage method with the appropriate parameters of the chosen algorithm.

We tested our strategy on the EURO STOXX 50 index, obtaining a partition into the $K=4$ cluster which offers a meaningful interpretation, as detailed in Sect. \ref{sec4}. Moreover, in Sect. \ref{subsec41}, we proposed an exogenous validation through portfolio diversification. By comparing a simple diversification strategy based on the clusters found by evidence accumulation with several competitors, compelling results are found with respect to different classical metrics of portfolio performance.

\newpage

\bibliography{bibliography}

\begin{thebibliography}{47}
\providecommand{\natexlab}[1]{#1}
\providecommand{\url}[1]{{#1}}
\providecommand{\urlprefix}{URL }
\providecommand{\doi}[1]{\url{https://doi.org/#1}}
\providecommand{\eprint}[2][]{\url{#2}}
 \bibcommenthead

\bibitem[{Alexander(2009)}]{alexander2009market}
Alexander C (2009) Market Risk Analysis, Boxset. John Wiley \& Sons

\bibitem[{Aroussi(2023)}]{aroussi2023reliably}
Aroussi R (2023) Reliably download historical market data from with python. Ran
  Aroussi

\bibitem[{Benevento et~al.(2024)Benevento, Durante, and
  Pappad{\`a}}]{benevento2024tail}
Benevento A, Durante F, Pappad{\`a} R (2024) Tail-dependence clustering of time
  series with spatial constraints. Environmental and Ecological Statistics
  31(3):801--817

\bibitem[{Bonanno et~al.(2004)Bonanno, Caldarelli, Lillo, Micciche, Vandewalle,
  and Mantegna}]{bonanno2004networks}
Bonanno G, Caldarelli G, Lillo F, et~al (2004) Networks of equities in
  financial markets. The European Physical Journal B 38(2):363--371

\bibitem[{Box et~al.(2015)Box, Jenkins, Reinsel, and Ljung}]{box2015time}
Box GE, Jenkins GM, Reinsel GC, et~al (2015) Time series analysis: forecasting
  and control. John Wiley \& Sons

\bibitem[{Charpentier and Segers(2007)}]{charpentier2007lower}
Charpentier A, Segers J (2007) Lower tail dependence for archimedean copulas:
  characterizations and pitfalls. Insurance: Mathematics and Economics
  40(3):525--532

\bibitem[{Charpentier and Segers(2009)}]{charpentier2009tails}
Charpentier A, Segers J (2009) Tails of multivariate archimedean copulas.
  Journal of Multivariate Analysis 100(7):1521--1537

\bibitem[{Chen and Fan(2006)}]{chen2006estimation}
Chen X, Fan Y (2006) Estimation and model selection of semiparametric
  copula-based multivariate dynamic models under copula misspecification.
  Journal of econometrics 135(1-2):125--154

\bibitem[{De~Luca and Zuccolotto(2011)}]{de2011tail}
De~Luca G, Zuccolotto P (2011) A tail dependence-based dissimilarity measure
  for financial time series clustering. Advances in data analysis and
  classification 5(4):323--340

\bibitem[{De~Luca and Zuccolotto(2021)}]{de2021hierarchical}
De~Luca G, Zuccolotto P (2021) Hierarchical time series clustering on tail
  dependence with linkage based on a multivariate copula approach.
  International Journal of Approximate Reasoning 139:88--103

\bibitem[{DeMiguel et~al.(2009)DeMiguel, Garlappi, and
  Uppal}]{demiguel2009optimal}
DeMiguel V, Garlappi L, Uppal R (2009) Optimal versus naive diversification:
  How inefficient is the 1/n portfolio strategy? The review of Financial
  studies 22(5):1915--1953

\bibitem[{Di~Lascio et~al.(2017)Di~Lascio, Durante, and Pappada}]{di2017copula}
Di~Lascio FML, Durante F, Pappada R (2017) Copula--based clustering methods.
  In: Copulas and Dependence Models with Applications: Contributions in Honor
  of Roger B. Nelsen. Springer, p 49--67

\bibitem[{Durante and Pappad{\`a}(2024)}]{durante2024dissimilarity}
Durante F, Pappad{\`a} R (2024) Dissimilarity-based clustering with soft
  proximity constraints. In: International Conference on Soft Methods in
  Probability and Statistics, Springer, pp 118--125

\bibitem[{Durante et~al.(2014)Durante, Pappad{\`a}, and
  Torelli}]{durante2014clustering}
Durante F, Pappad{\`a} R, Torelli N (2014) Clustering of financial time series
  in risky scenarios. Advances in Data Analysis and Classification
  8(4):359--376

\bibitem[{Durante et~al.(2015{\natexlab{a}})Durante, Fernandez-Sanchez, and
  Pappada}]{durante2015copulas}
Durante F, Fernandez-Sanchez J, Pappada R (2015{\natexlab{a}}) Copulas,
  diagonals, and tail dependence. Fuzzy Sets and Systems 264:22--41

\bibitem[{Durante et~al.(2015{\natexlab{b}})Durante, Pappad{\`a}, and
  Torelli}]{durante2015clustering}
Durante F, Pappad{\`a} R, Torelli N (2015{\natexlab{b}}) Clustering of time
  series via non-parametric tail dependence estimation. Statistical Papers
  56(3):701--721

\bibitem[{Durante et~al.(2016)Durante, Sempi et~al.}]{durante2016principles}
Durante F, Sempi C, et~al (2016) Principles of copula theory, vol 474. CRC
  press Boca Raton, FL

\bibitem[{D’Urso et~al.(2013)D’Urso, Di~Lallo, and
  Maharaj}]{d2013autoregressive}
D’Urso P, Di~Lallo D, Maharaj EA (2013) Autoregressive model-based fuzzy
  clustering and its application for detecting information redundancy in air
  pollution monitoring networks. Soft Computing 17(1):83--131

\bibitem[{D’Urso et~al.(2023)D’Urso, De~Luca, Vitale, and
  Zuccolotto}]{d2023tail}
D’Urso P, De~Luca G, Vitale V, et~al (2023) Tail dependence-based fuzzy
  clustering of financial time series. Annals of Operations Research pp 1--27

\bibitem[{Elton et~al.(2009)Elton, Gruber, Brown, and
  Goetzmann}]{elton2009modern}
Elton EJ, Gruber MJ, Brown SJ, et~al (2009) Modern portfolio theory and
  investment analysis. John Wiley \& Sons

\bibitem[{Everitt et~al.(2011)Everitt, Landau, Leese, Stahl
  et~al.}]{everitt2011hierarchical}
Everitt BS, Landau S, Leese M, et~al (2011) Hierarchical clustering. Cluster
  analysis 5(71-110):71--110

\bibitem[{Fred and Jain(2002)}]{fred2002data}
Fred AL, Jain AK (2002) Data clustering using evidence accumulation. In: 2002
  International conference on pattern recognition, IEEE, pp 276--280

\bibitem[{Fuchs et~al.(2021)Fuchs, Di~Lascio, and
  Durante}]{fuchs2021dissimilarity}
Fuchs S, Di~Lascio FML, Durante F (2021) Dissimilarity functions for
  rank-invariant hierarchical clustering of continuous variables. Computational
  Statistics \& Data Analysis 159:107201

\bibitem[{Glosten et~al.(1993)Glosten, Jagannathan, and
  Runkle}]{glosten1993relation}
Glosten LR, Jagannathan R, Runkle DE (1993) On the relation between the
  expected value and the volatility of the nominal excess return on stocks. The
  journal of finance 48(5):1779--1801

\bibitem[{Gudendorf and Segers(2010)}]{gudendorf2010extreme}
Gudendorf G, Segers J (2010) Extreme-value copulas. In: Copula Theory and Its
  Applications: Proceedings of the Workshop Held in Warsaw, 25-26 September
  2009, Springer, pp 127--145

\bibitem[{Haugen and Baker(1991)}]{haugen1991efficient}
Haugen RA, Baker NL (1991) The efficient market inefficiency of
  capitalization-weighted stock portfolios. Journal of portfolio management
  17(3):35

\bibitem[{Hubert and Arabie(1985)}]{hubert1985comparing}
Hubert L, Arabie P (1985) Comparing partitions. Journal of classification
  2(1):193--218

\bibitem[{Joe(1993)}]{joe1993parametric}
Joe H (1993) Parametric families of multivariate distributions with given
  margins. Journal of multivariate analysis 46(2):262--282

\bibitem[{Joe(1997)}]{joe1997multivariate}
Joe H (1997) Multivariate models and multivariate dependence concepts. CRC
  press

\bibitem[{Joe and Xu(1996)}]{joe1996estimation}
Joe H, Xu JJ (1996) The estimation method of inference functions for margins
  for multivariate models. Faculty Research and Publications

\bibitem[{Liao(2005)}]{liao2005clustering}
Liao TW (2005) Clustering of time series data—a survey. Pattern recognition
  38(11):1857--1874

\bibitem[{Maharaj et~al.(2019)Maharaj, D'Urso, and Caiado}]{maharaj2019time}
Maharaj EA, D'Urso P, Caiado J (2019) Time series clustering and
  classification. Chapman and Hall/CRC

\bibitem[{Mantegna(1999)}]{mantegna1999hierarchical}
Mantegna RN (1999) Hierarchical structure in financial markets. The European
  Physical Journal B-Condensed Matter and Complex Systems 11(1):193--197

\bibitem[{Monti et~al.(2003)Monti, Tamayo, Mesirov, and
  Golub}]{monti2003consensus}
Monti S, Tamayo P, Mesirov J, et~al (2003) Consensus clustering: a
  resampling-based method for class discovery and visualization of gene
  expression microarray data. Machine learning 52(1):91--118

\bibitem[{Nicolas(2022)}]{nicolas2022pycop}
Nicolas ML (2022) pycop: a python package for dependence modeling with copulas.
  Zenodo Software Package 70:7030034

\bibitem[{Patton(2009)}]{patton2009copula}
Patton AJ (2009) Copula--based models for financial time series. In: Handbook
  of financial time series. Springer, p 767--785

\bibitem[{Patton(2012)}]{patton2012review}
Patton AJ (2012) A review of copula models for economic time series. Journal of
  Multivariate Analysis 110:4--18

\bibitem[{Piccolo(1990)}]{piccolo1990distance}
Piccolo D (1990) A distance measure for classifying arima models. Journal of
  time series analysis 11(2):153--164

\bibitem[{R{\'e}millard(2017)}]{remillard2017goodness}
R{\'e}millard B (2017) Goodness-of-fit tests for copulas of multivariate time
  series. Econometrics 5(1):13

\bibitem[{Rockafellar et~al.(2000)Rockafellar, Uryasev
  et~al.}]{rockafellar2000optimization}
Rockafellar RT, Uryasev S, et~al (2000) Optimization of conditional
  value-at-risk. Journal of risk 2:21--42

\bibitem[{Rousseeuw(1987)}]{rousseeuw1987silhouettes}
Rousseeuw PJ (1987) Silhouettes: a graphical aid to the interpretation and
  validation of cluster analysis. Journal of computational and applied
  mathematics 20:53--65

\bibitem[{Schmid et~al.(2010)Schmid, Schmidt, Blumentritt, Gai{\ss}er, and
  Ruppert}]{schmid2010copula}
Schmid F, Schmidt R, Blumentritt T, et~al (2010) Copula-based measures of
  multivariate association. In: Copula Theory and Its Applications: Proceedings
  of the Workshop Held in Warsaw, 25-26 September 2009, Springer, pp 209--236

\bibitem[{Sibuya et~al.(1960)}]{sibuya1960bivariate}
Sibuya M, et~al (1960) Bivariate extreme statistics. Annals of the Institute of
  Statistical Mathematics 11(2):195--210

\bibitem[{Sklar(1959)}]{sklar1959fonctions}
Sklar M (1959) Fonctions de r{\'e}partition {\`a} n dimensions et leurs marges.
  In: Annales de l'ISUP, pp 229--231

\bibitem[{Sweeting and Fotiou(2013)}]{sweeting2013calculating}
Sweeting P, Fotiou F (2013) Calculating and communicating tail association and
  the risk of extreme loss. British Actuarial Journal 18(1):13--72

\bibitem[{Tsay(2005)}]{tsay2005analysis}
Tsay RS (2005) Analysis of financial time series. John wiley \& sons

\bibitem[{Wang et~al.(2016)Wang, Pappad{\`a}, Durante, and
  Foscolo}]{wang2016portfolio}
Wang H, Pappad{\`a} R, Durante F, et~al (2016) A portfolio diversification
  strategy via tail dependence clustering. In: Soft methods for data science.
  Springer, p 511--518

\end{thebibliography}

\newpage

\begin{appendices}

\section{Marginal modeling of time series}\label{secA1}

The log-returns of the $d = 38$ selected time series are preliminarily processed to account for their possible autocorrelation and heteroschedasticity. Details about such results are provided in this appendix Section. 

First, distinct ARMA models are fit to each time series, with the orders constrained not to exceed $5$ for both their autoregressive and moving average components; further details on ARMA models and their notation can be found in \cite{box2015time}. We report in Table \ref{tabA1} the intercepts and the coefficients of the autoregressive components of the ARMA models and in Table \ref{tabA2} those of their moving average components, as well as the residuals variance. Values are rounded to $4$ decimal places; the omitted coefficients are relative to the components excluded by model selection, according to the Akaike information criterion. The statistical significance of all parameters was assessed by inspecting their $p$-values.

\begin{table}[h]
\caption{ARMA models intercepts and autoregressive coefficients. Ticker symbols are used for brevity; we refer to Table \ref{tab1} for the full correspondences}\label{tabA1}%
\begin{tabular}{@{}lllllll@{}}
\toprule
Ticker symbol & c & $\phi_1$ & $\phi_2$ & $\phi_3$ & $\phi_4$ & $\phi_5$ \\
\midrule
ADS.DE &  &  &  &  &  & \\
AD.AS &  &  &  &  &  & \\
AI.PA &  & -0.4666 &  &  &  & \\
AIR.PA &  & -1.0724 & -0.9647 & -0.6108 &  & \\
ALV.DE &  &  &  &  &  & \\
ABI.BR &  &  &  &  & \\
ASML.AS & 0.0011 &  &  &  &  & \\
CS.PA &  & 0.0343 & 0.1473 & 0.0501 &  & \\
BAS.DE &  &  &  &  &  & \\
BAYN.DE &  &  &  &  &  & \\
SAN.MC &  & -1.0198 & -0.7394 &  &  & \\
BMW.DE &  &  &  &  &  & \\
BNP.PA &  &  &  &  &  & \\
BN.PA &  &  &  &  &  & \\
DHL.DE &  &  &  &  &  & \\
DTE.DE &  & 0.8082 & -0.3528 &  &  & \\
ENEL.MI &  &  &  &  &  & \\
ENI.MI &  &  &  &  &  & \\
EL.PA &  &  &  &  &  & \\
FLTR.L &  & -0.3463 & -0.6113 &  &  & \\
RMS.PA & 0.0009 &  &  &  &  & \\
IBE.MC & 0.0006 & -0.0601 & 0.0549 &  &  & \\
ITX.MC &  &  &  &  &  &\\
IFX.DE &  &  &  &  &  & \\
INGA.AS &  & 0.7677 &  &  &  & \\
ISP.MI &  & 0.7121 &  &  &  & \\
OR.PA &  & -0.0655 &  &  &  & \\
MC.PA & 0.0009 & -0.0461 &  &  &  & \\
MBG.DE &  & -1.4513 & -0.5815 & 0.1099 &  & \\
MUV2.DE &  &  &  &  &  & \\
SAF.PA &  &  &  &  &  & \\
SAN.PA &  &  &  &  &  & \\
SAP.DE &  &  &  &  &  & \\
SU.PA &  &  &  &  &  & \\
SIE.DE &  &  &  &  &  & \\
TTE.PA &  &  &  &  &  & \\
DG.PA &  & -0.0946 &  &  &  & \\
VOW.DE &  &  &  &  &  & \\
\botrule
\end{tabular}
\end{table}

\begin{table}[h]
\caption{ARMA models moving average coefficients and residuals variances. Ticker symbols are used for brevity; we refer to Table \ref{tab1} for the full correspondences}\label{tabA2}%
\begin{tabular}{@{}lllllll@{}}
\toprule
Ticker symbol & $\theta_1$ & $\theta_2$ & $\theta_3$ & $\theta_4$ & $\theta_5$ & $\sigma^2$ \\
\midrule
ADS.DE &  &  &  &  &  & 0.0005 \\
AD.AS &  &  &  &  &  & 0.0002 \\
AI.PA & 0.3397 &  &  &  &  & 0.0002 \\
AIR.PA & 1.1335 & 0.9791 & 0.6322 & 0.0704 & 0.1212 & 0.0007 \\
ALV.DE &  &  &  &  &  & 0.0003 \\
ABI.BR &  &  &  &  &  & 0.0004 \\
ASML.AS &  &  &  &  &  & 0.0005 \\
CS.PA &  &  &  &  &  & 0.0003 \\
BAS.DE &  &  &  &  &  & 0.0004 \\
BAYN.DE &  &  &  &  &  & 0.0004 \\
SAN.MC & 1.0364 & 0.8266 &  &  &  & 0.0005 \\
BMW.DE &  &  &  &  &  & 0.0004 \\
BNP.PA &  &  &  &  &  & 0.0005 \\
BN.PA &  &  &  &  &  & 0.0002 \\
DHL.DE &  &  &  &  &  & 0.0003 \\
DTE.DE & -0.8795 & 0.4981 &  &  &  & 0.0002 \\
ENEL.MI &  &  &  &  &  & 0.0003 \\
ENI.MI &  &  &  &  &  & 0.0004 \\
EL.PA &  &  &  &  &  & 0.0003 \\
FLTR.L & 0.3931 & 0.7105 & 0.0032 & 0.1222 &  & 0.0006 \\
RMS.PA &  &  &  &  &  & 0.0003 \\
IBE.MC &  &  &  &  &  & 0.0002 \\
ITX.MC &  &  &  &  &  & 0.0004 \\
IFX.DE &  &  &  &  &  & 0.0006 \\
INGA.AS & -0.7038 &  &  &  &  & 0.0006 \\
ISP.MI & -0.7233 & 0.0769 &  &  &  & 0.0004 \\
OR.PA &  &  &  &  &  & 0.0002 \\
MC.PA &  &  &  &  &  & 0.0003 \\
MBG.DE & 1.4642 & 0.6823 &  &  &  & 0.0005 \\
MUV2.DE &  &  &  &  &  & 0.0003 \\
SAF.PA &  &  &  &  &  & 0.0007 \\
SAN.PA &  &  &  &  &  & 0.0002 \\
SAP.DE &  &  &  &  &  & 0.0003 \\
SU.PA &  &  &  &  &  & 0.0003 \\
SIE.DE &  &  &  &  &  & 0.0004 \\
TTE.PA & 0.0578 &  &  &  &  & 0.0004 \\
DG.PA &  &  &  &  &  & 0.0004 \\
VOW.DE &  &  &  &  &  & 0.0006 \\
\botrule
\end{tabular}
\end{table}

A GJR-GARCH$(1, 1)$ model with Student's $t$ distributed errors is separately fitted to each time series of residuals of the ARMA models. Such GARCH models are selected to capture asymmetries in the volatility processes; see \cite{glosten1993relation} for further information and notation details. Distinct constant mean processes are assumed for each model. Table \ref{tabA3} reports the estimated parameters and coefficients from the fitted models, rounded to $4$ decimal places. As before, inspecting the $p$-values of such parameters allowed to assess their statistical significance. The series of standardized residuals $\{\hat{\varepsilon}_{it}\}$, for $i = 1, \dots, d$ and $t = 1, \dots, T$, obtained from such models are used as input for the subsequent copula-based dependence modeling.

\begin{table}[h]
\caption{GJR-GARCH$(1, 1)$ models parameters and coefficients. Ticker symbols are used for brevity; we refer to Table \ref{tab1} for the full correspondences}\label{tabA3}%
\begin{tabular}{@{}lllllll@{}}
\toprule
Ticker symbol & $\mu$ & $\omega$ & $q$ & $\gamma$ & $\beta$ & $\nu$ \\
\midrule
ADS.DE & -0.0043 & 0.1048 & 0.0048 & 0.0833 & 0.9355 & 4.0147 \\
AD.AS & 0.0607 & 0.4344 & 0.1443 & 0.0422 & 0.5903 & 4.1499 \\
AI.PA & 0.0615 & 0.0612 & 0.0131 & 0.1272 & 0.8846 & 6.2299 \\
AIR.PA & 0.0050 & 0.1665 & 0.0065 & 0.2084 & 0.8702 & 5.5291 \\
ALV.DE & 0.0696 & 0.0914 & 0.0180 & 0.2164 & 0.8431 & 3.8817 \\
ABI.BR & -0.0340 & 0.0410 & 0.0157 & 0.0285 & 0.9611 & 3.3662 \\
ASML.AS & 0.0438 & 0.1311 & 0.0028 & 0.1023 & 0.9193 & 6.2184 \\
CS.PA & 0.0373 & 0.0290 & 0.0000 & 0.1287 & 0.9257 & 4.5984 \\
BAS.DE & -0.0397 & 0.0278 & 0.0000 & 0.0639 & 0.9585 & 4.8156 \\
BAYN.DE & -0.0418 & 0.4706 & 0.0013 & 0.1670 & 0.8107 & 3.5635 \\
SAN.MC & -0.0518 & 0.0549 & 0.0212 & 0.0657 & 0.9356 & 7.7411 \\
BMW.DE & -0.0024 & 0.0775 & 0.0126 & 0.0886 & 0.9200 & 4.7557 \\
BNP.PA & 0.0025 & 0.0310 & 0.0021 & 0.1044 & 0.9396 & 6.3827 \\
BN.PA & 0.0144 & 0.0317 & 0.0115 & 0.0565 & 0.9407 & 4.3345 \\
DHL.DE & 0.0382 & 0.0664 & 0.0000 & 0.1030 & 0.9248 & 5.3524 \\
DTE.DE & 0.0141 & 0.0299 & 0.0181 & 0.0693 & 0.9310 & 4.4999 \\
ENEL.MI & 0.0572 & 0.1141 & 0.0147 & 0.1307 & 0.8707 & 5.0321 \\
ENI.MI & 0.0707 & 0.0592 & 0.0249 & 0.0749 & 0.9138 & 5.8395 \\
EL.PA & 0.0773 & 0.0420 & 0.0246 & 0.1071 & 0.9125 & 4.4891 \\
FLTR.L & -0.0404 & 0.1013 & 0.0101 & 0.0854 & 0.9362 & 3.9027 \\
RMS.PA & 0.0845 & 0.0203 & 0.0223 & 0.0619 & 0.9406 & 5.1797 \\
IBE.MC & -0.0102 & 0.0755 & 0.0815 & 0.1229 & 0.8273 & 5.6679 \\
ITX.MC & 0.0115 & 0.0621 & 0.0148 & 0.0597 & 0.9393 & 4.8790 \\
IFX.DE & 0.0015 & 0.1838 & 0.0000 & 0.0895 & 0.9242 & 8.1751 \\
INGA.AS & -0.0279 & 0.0406 & 0.0000 & 0.1237 & 0.9356 & 4.5061 \\
ISP.MI & 0.0274 & 0.0879 & 0.0000 & 0.1715 & 0.8957 & 5.2571 \\
OR.PA & 0.0831 & 0.0357 & 0.0000 & 0.1025 & 0.9334 & 5.5742 \\
MC.PA & 0.0277 & 0.1102 & 0.0000 & 0.1392 & 0.9014 & 5.0672 \\
MBG.DE & 0.0006 & 0.0629 & 0.0000 & 0.0940 & 0.9418 & 5.1209 \\
MUV2.DE & 0.1174 & 0.0162 & 0.0000 & 0.0690 & 0.9579 & 3.8179 \\
SAF.PA & 0.0247 & 0.1085 & 0.0151 & 0.2683 & 0.8507 & 5.3174 \\
SAN.PA & 0.0450 & 0.1691 & 0.0273 & 0.1125 & 0.8226 & 6.8261 \\
SAP.DE & 0.0570 & 0.2773 & 0.0000 & 0.1180 & 0.8440 & 3.8768 \\
SU.PA & 0.0824 & 0.0755 & 0.0000 & 0.1100 & 0.9185 & 7.2097 \\
SIE.DE & 0.0376 & 0.0491 & 0.0066 & 0.0674 & 0.9447 & 5.0867 \\
TTE.PA & 0.0555 & 0.0458 & 0.0236 & 0.0763 & 0.9225 & 6.3984 \\
DG.PA & 0.0178 & 0.1150 & 0.0050 & 0.2044 & 0.8513 & 5.3389 \\
VOW.DE & -0.0221 & 0.2490 & 0.0755 & 0.1040 & 0.8329 & 5.5394 \\
\botrule
\end{tabular}
\end{table}

\end{appendices}
\newpage

\end{document}